\documentclass[preprint,showpacs,preprintnumbers,amsmath,amssymb]{revtex4}
\usepackage{graphicx}

\def\be{\begin{equation}}
\def\ee{\end{equation}}
\def\bee{\begin{eqnarray}}
\def\eee{\end{eqnarray}}

\begin{document}
\title{Ion temperature gradient instability at sub-Larmor radius scales with non-zero ballooning angle}

\author{P. Migliano$^1$, Y. Camenen$^2$, F.J. Casson$^3$, W.A. Hornsby$^1$, A.G. Peeters$^1$}

\vskip 0.2 truecm 

\address{$^1$ University of Bayreuth, Physics department, Universit\" atsstrasse 30 Bayreuth, Germany}
\address{$^2$ Aix-Marseille Universit\'{e}, CNRS, PIIM UMR 7345, 13397 Marseille, France}
\address{$^3$ Max Planck Institut fuer Plasmaphysik, EURATOM association, Boltzmannstrasse 2, 85748 Garching, Germany}


\begin{abstract}
Linear gyro-kinetic stability calculations predict unstable toroidal Ion Temperature Gradient modes with 
normalised poloidal wave vectors well above one ($k_\theta \rho_i > 1$) for standard parameters and with adiabatic 
electrons. 
These modes have a maximum amplitude at a poloidal angle $\theta$ that is shifted away from the low field side
($\theta \ne 0$). 
The physical mechanism is clarified through the use of a fluid model.
It is shown that the shift of the mode away from the low field side ($\theta \ne 0$) reduces the effective drift 
frequency, and allows for the instability to develop. 
Numerical tests using the gyro-kinetic model confirm this physical mechanism. 
It is furthermore shown that modes with $\theta \ne 0$ can be important also for $k_\theta \rho_i < 1$ close 
to the threshold of the ITG. 
In fact, modes with $\theta \ne 0$ can exist for normalised temperature gradient lengths below the threshold of 
the ITG obtained for $\theta = 0$. 
\end{abstract}

\maketitle

\section{INTRODUCTION} 

The growth rate of the ion temperature gradient mode (ITG) as a function of the 
normalised poloidal wave vector $k_\theta \rho_i$, has been reported many times 
in the literature, see for instance \cite{DIM00}, as a bell shaped curve with a 
single maximum.
Here, $k_\theta$ is the poloidal component of the wave vector and $\rho_i
= m_i v_{thi} / Z e B = \sqrt{2 m_i T_i}/ZeB$ is the ion Larmor radius, 
with $m_i$ the ion mass, $v_{thi}$ the ion thermal velocity, $Z$ the charge
number, $e$ the elementary charge, $B$ the magnetic field strength, and 
$T_i$ the ion temperature. 

In this paper we report on collisionless Ion Temperature Gradient (ITG) instabilities
with $k_\theta\rho_i > 1$ and adiabatic electrons.
It will be shown that these instabilities can exist for relevant Tokamak 
parameters. 
The physical mechanism of these instabilities will be shown to be related to 
the reduction of the effective drift frequency through the shift of the mode away 
from the low field side position. 
This mechanism is different from previously reported \cite{SMO02,HIR02,GAO03,GAO05,CHO09}
instabilities with $k_\theta \rho_i > 1$, which are unstable only in a slab or for weak 
toroidicity. 

This paper is structured as follows: Section II introduces the high $k_\theta$ ITG
through numerical simulation based on the gyro-kinetic model, Section III discusses
the physics of the instability through the use of a simple fluid model. Section 
IV discusses the relation with previously published work and, finally Section V gives 
the conclusions.  

\section{HIGH $k_\theta$ ITG} 
 
An example of a calculation using the ballooning transform \cite{CON78} (the use of this 
transform is referred to below as the 'spectral case'), obtained with the gyrokinetic code GKW \cite{pee09b}, 
is given in Fig.~1 by the dash-dotted curve that has a single maximum.
The parameters of this, and all other simulations in this paper, are those of the 
Waltz standard case \cite{WAL95}: ion temperature gradient lengths $R/L_{Ti}=9.0$ 
density gradient length $R/L_{ni}=3.0$, electron and ion temperature $T_e = 
T_i$, safety factor $q = 2$, magnetic shear $\hat s = 1$, and inverse aspect ratio
$\epsilon = 0.166$. The simulations use circular geometry retaining finite $\epsilon$ 
effects, and the flux tube approximation is always applied. 

However, GKW simulations with the radial 
direction described using finite
 differences (simulations that use finite difference in 
the radial direction are referred to below as the 'non-spectral case') show a
surprisingly different behaviour for $k_\theta\rho_i > 0.6$, displaying a
spectrum with two maxima and having unstable modes with $k_\theta \rho_i$ well above 
one, as is shown by the full line of Fig.~1.

\begin{figure}[h]
 \begin{center}
  \includegraphics[width=9cm]{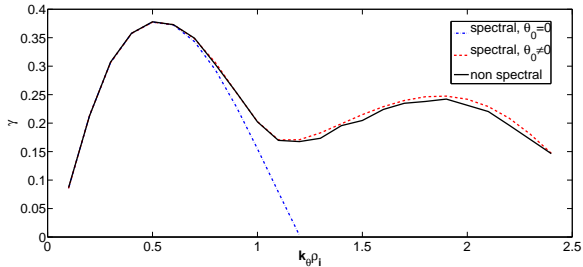}
  \caption{\small (Colour on-line) Growth rate ($\gamma$ in units $v_{thi}/R_0$ where $R_0$ is the major radius 
of the magnetic axis) as a function of $k_\theta\rho_i$.  The dash-dotted (blue) line is the spectral case with 
$\theta_0=0$, the full (black) line is the non-spectral case, and the dashed (red) line is the maximum growth rated obtained 
when varying $\theta_0$ in the spectral case.}
  \label{gamma_vs_kthrho_spectral_chin_00_ne00_global}
 \end{center}
\end{figure}



\begin{figure}[h]
 \begin{center}
  \includegraphics[width=9cm]{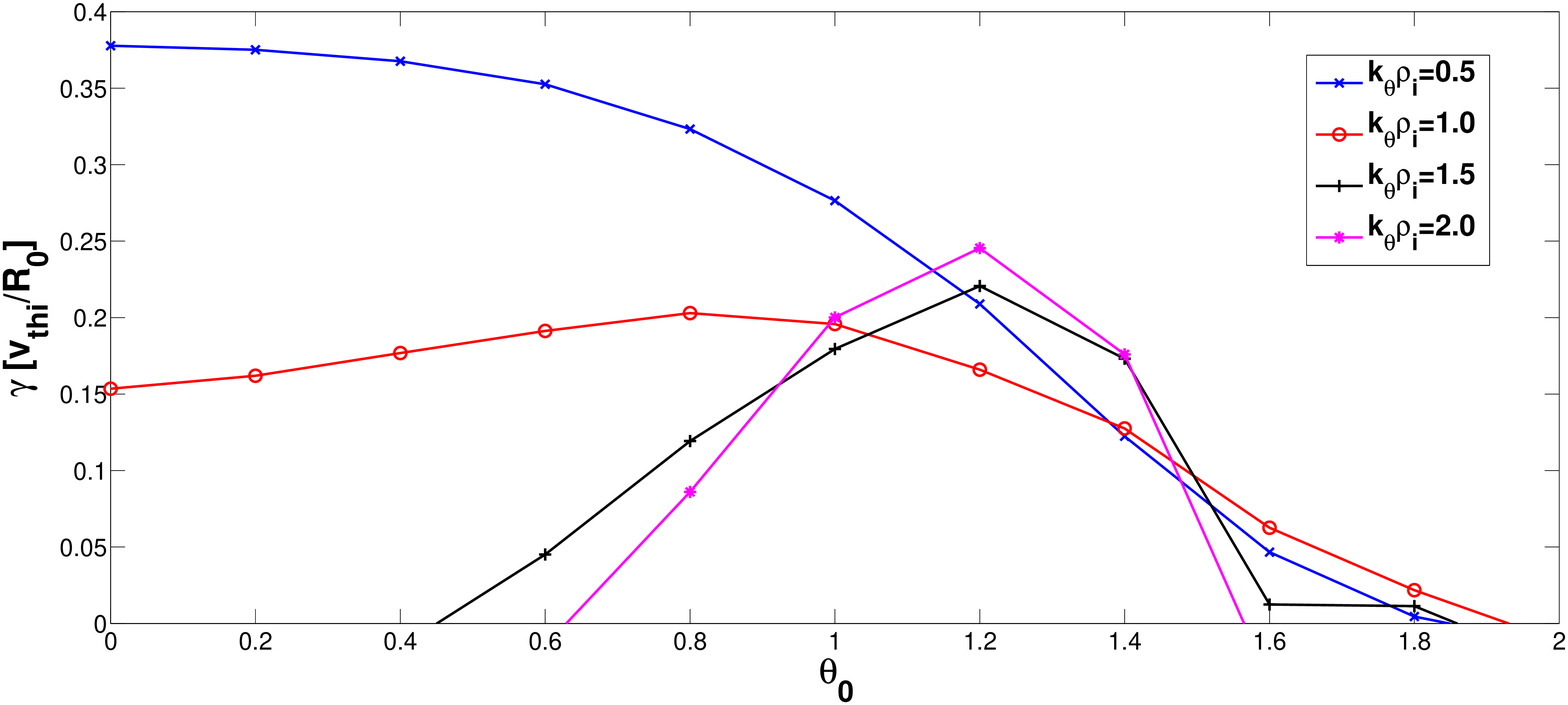}
  \caption{\small (Colour on-line) Growth rates of the spectral case ($\gamma$ in units $v_{thi}/R_0$) as a function of $\theta_0$ for four 
representative values of $k_\theta\rho_i$. The curves are denoted as follows: $k_\theta \rho_i = 0.5$ (blue) crosses, $k_\theta \rho_i = 1.0$ (red)
circles, $k_\theta \rho_i = 1.5$ (black) pluses, $k_\theta \rho_i = 2.0$ (magenta) stars}
  \label{gamma_vs_chin_spectral_kthrho_scan}
 \end{center}
\end{figure}

\begin{figure}[h]
 \begin{center}
  \includegraphics[width=9cm]{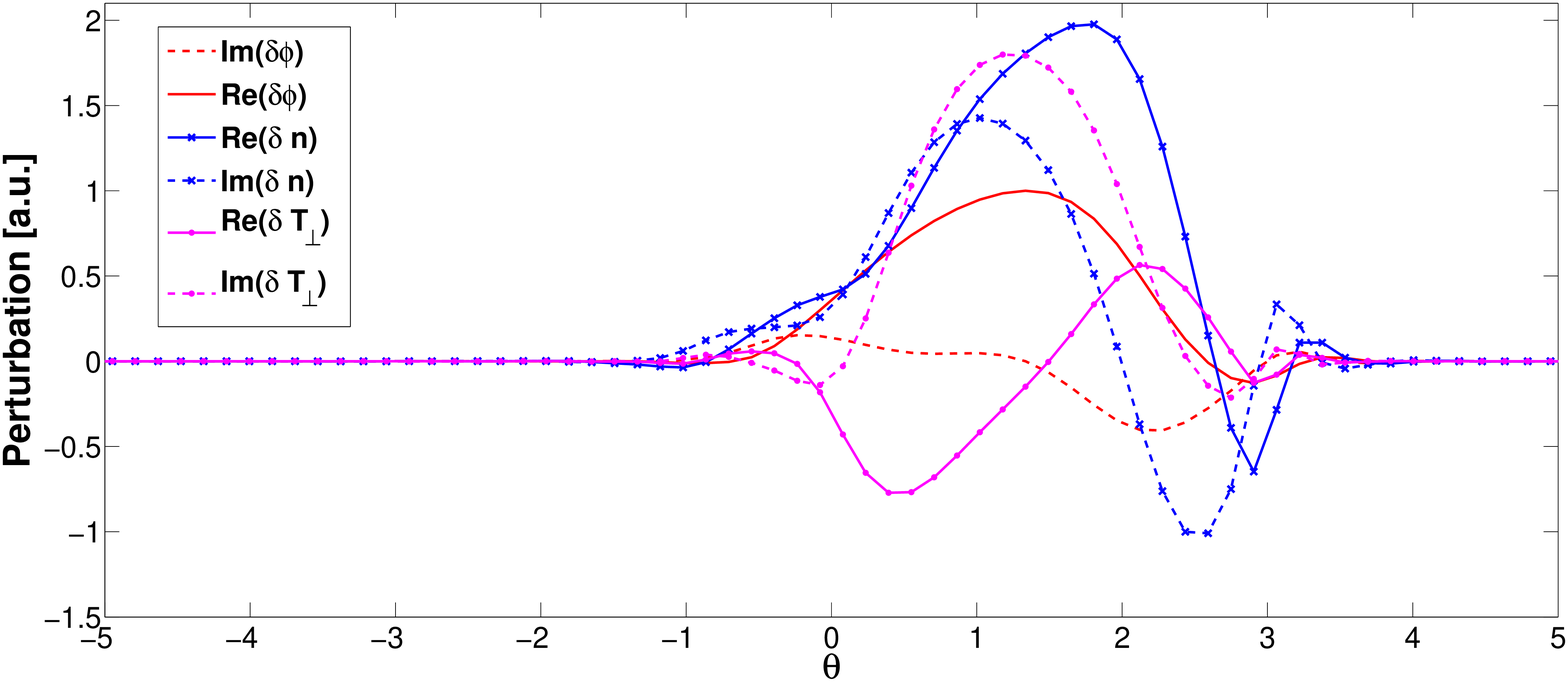}
  \caption{\small (Colour on-line) Eigenfunction in arbitrary units as a function of the poloidal angle $\theta$. Full lines give the real, whereas dashed lines give the imaginary part. The (red) lines without symbols is the potential perturbation ($\delta\phi$), the (blue) lines with the symbol 'x' the density perturbation ($\delta n$), and the (magenta) lines with the closed circles is the perpendicular temperature perturbation ($\delta T_\perp$).}
\label{mode_structure} \end{center}
\end{figure}

The essential difference between these two simulations is the number of radial modes that are kept in each of them. 
There are many more radial modes in the non-spectral case compared to the spectral one, in which the radial wave vector 
($k_r$) is set by the condition of the field alignment of the mode \cite{CON78}
\begin{equation}
 k_r = \hat{s}\theta k_\theta \ ,
 \label{radial_component_wave_vector_chin00}
\end{equation}
and is zero at the low field side position (poloidal angle $\theta = 0$). This suggests that the unstable modes for $k_\theta \rho_i > 1$ 
have a finite radial wave vector at the low field side position.  
It is well known, that a finite radial wave vector can be introduced in the ballooning transform through the introduction of the angle 
$\theta_0$ such that 
 \begin{equation}
 k_r = \hat{s}(\theta-\theta_0) k_\theta \ ,
 \label{radial_component_wave_vector_chin_ne00}
\end{equation} 
The growth rate as a function of $\theta_0$ for various values of $k_\theta \rho_i$ is shown in 
Fig.~\ref{gamma_vs_chin_spectral_kthrho_scan}. 
For $k_\theta\rho_i > 0.6$ the most unstable mode has a finite $\theta_0$ and the mode is shifted away from the low field side, as shown in Fig.~\ref{mode_structure}, which displays the eigen function along the magnetic 
field for $k_\theta \rho_i = 1.5$ and $\theta_0 = 1.2$. There is no preferred sign for $\theta_0$ and the mode with $\theta_0=-1.2$ is equally unstable but shifted in the negative $\theta$ direction.
Taking the maximum growth rate (by varying $\theta_0$) for each $k_\theta \rho_i$ from 
the spectral simulations yields the dashed (red) curve in Fig.~\ref{gamma_vs_kthrho_spectral_chin_00_ne00_global}. 
There is agreement between the spectral and the non-spectral cases, as there should be. 

\begin{figure}[h]
 \begin{center}
  \includegraphics[width=9cm]{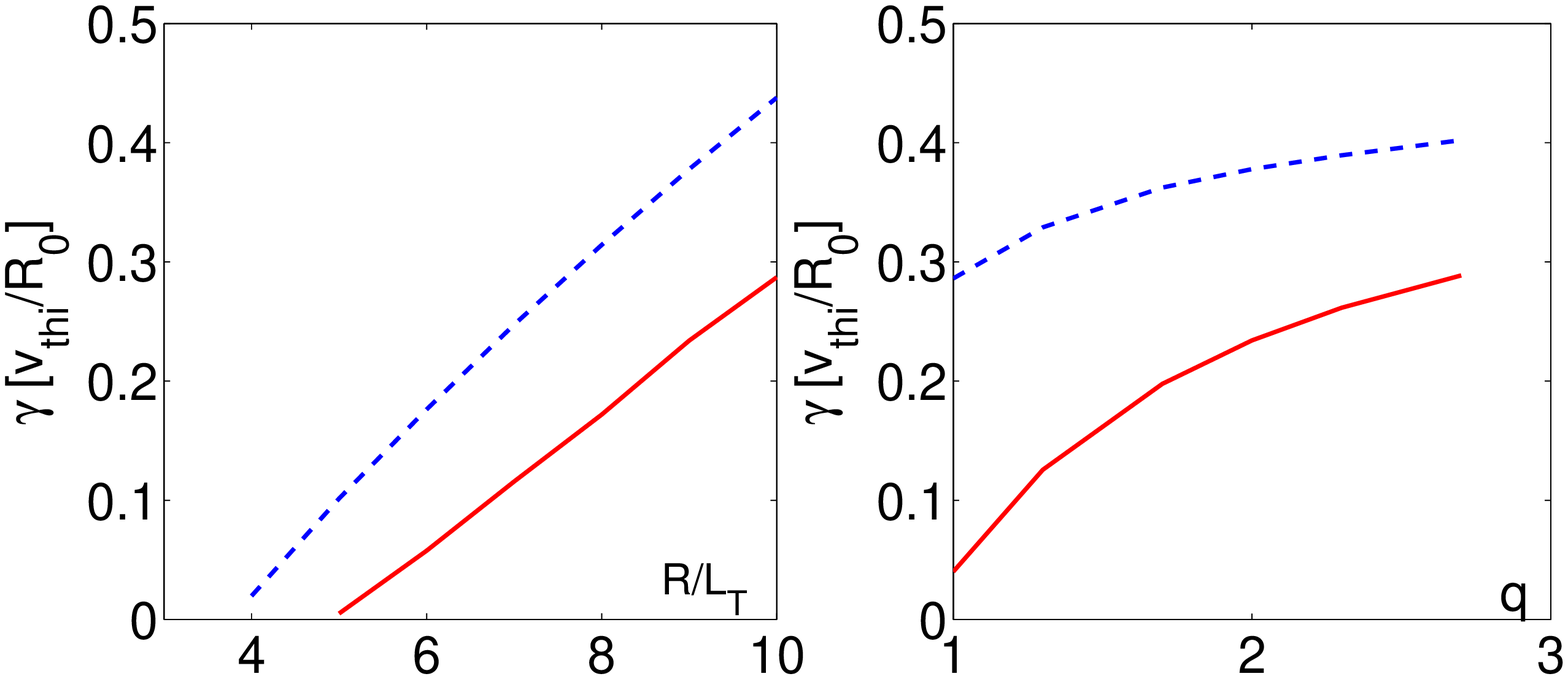}
  \caption{\small (Colour on-line) The growth rate as a function of $R/L_T$ (left) and the safety factor $q$ (right). The (red) full line 
gives the result for $k_\theta \rho_i = 1.9$, while the (blue) dashed line gives the result for $k_\theta \rho_i = 0.5$.
\label{rltqscan}}
 \end{center}
\end{figure}

\begin{figure}[h]
 \begin{center}
  \includegraphics[width=9cm]{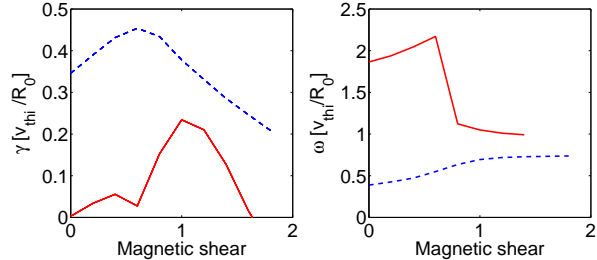}
  \caption{\small (Colour on-line) The growth rate (left) and frequency (right) as a function of the magnetic shear. 
The (red) full line gives the result for $k_\theta \rho_i = 1.9$, while the (blue) dashed line gives the result for 
$k_\theta \rho_i = 0.5$.
\label{shatscan}}
 \end{center}
\end{figure}

In order to get further insight into the high $k_\theta$ ITGs several parameter scans have been performed. 
These are done varying one parameter while keeping all the other parameters fixed to the standard case. The results 
of the $R/L_T$, $q$ and $\hat s$ scans are shown in Figs. \ref{rltqscan} and \ref{shatscan}. 
All calculations are performed using the non-spectral setup with $k_\theta \rho_i = 1.9$. For comparison, 
also the results for $k_\theta \rho_i = 0.5$ are shown. 	

Fig. \ref{rltqscan} shows that the growth rate of the high $k_\theta$ ITG increases with $R/L_T$ very similar 
to the $k_\theta \rho_i = 0.5$ mode. The mode has a higher threshold in $R/L_T$ though is more stable over 
the entire scan. In the same figure also the dependence on the safety factor is shown. A larger safety factor 
increases the connection length between the low and high field side and, therefore, works destabilising. This is 
the case for $k_\theta \rho_i = 0.5$, but to a much larger extent for the high $k_\theta$ ITG. 
The localisation of the mode at $\theta_0 \ne 0$ makes the requirement of a sufficient long field line length  
less easy to satisfy. 

Fig.~\ref{shatscan} gives the value of the growth rate as a function of the magnetic shear. The growth rate curve 
has two maxima for $k_\theta \rho_i = 1.9$, and it can be verified from the Figure of the frequency that these maxima 
belong to two different modes. 
A high growth rate is obtained only at sufficiently large shear. A high shear reduces the width of the eigenmode and is 
therefore beneficial for the $\theta_0 \ne 0$ modes. The dependence of the growth rate on the inverse aspect 
ratio $\epsilon$ (not shown) is found to be relatively weak.

\section{PHYSICAL MECHANISM}

An understanding of the physics of the high $k_\theta$ ITGs can be obtained by considering a simple fluid model. 
Here, the equations and normalisation given in \cite{pee09c} are used, and the reader is referred to this paper for 
details on the derivation. 
The gyro-kinetic equation, neglecting the parallel derivatives can be written in the form (see Eq.~(68) of Ref.~\cite{pee09c}): 
\begin{equation} 
{\partial f \over \partial t} + {\bf v}_D \cdot \nabla f = -{\bf v}_E \cdot \nabla_p F_M - {\bf v}_D \cdot {Z e 
\nabla \langle \phi \rangle \over T} F_M 
\label{gyrokineticequation}
\end{equation} 
where ${\bf v}_D$ is the drift due to the magnetic field inhomogeneity, ${\bf v}_E$ is the perturbed ExB velocity, 
$f$ the perturbed distribution, $\phi$ the perturbed electrostatic potential, and $\nabla_p$ is defined through 
Eq.~(69) of Ref.~\cite{pee09c}. In comparison to Ref.~\cite{pee09c} the plasma rotation will be neglected, but it 
will not be assumed that the mode is localised on the low field side. Assuming a concentric circular magnetic equilibrium, 
\begin{equation}
{\bf v}_D \cdot \nabla = {\rm i} v_D k_\theta \cos\theta + {\rm i} v_D k_r \sin\theta = {\rm i} k_\theta v_D K 
\end{equation} 
where $\theta$ is the poloidal angle, and $k_\theta$ ($k_r$) is the poloidal (radial) wave vector. In the equation 
above $K$ is introduced to shorten the mathematics. Using Eq.~(\ref{radial_component_wave_vector_chin_ne00}) one obtains
\begin{equation} 
K = \cos \theta + \hat s (\theta - \theta_0) \sin\theta 
\end{equation}  
$K$ measures the dependence of the convective derivative (${\bf v}_D \cdot \nabla$) on the poloidal angle. 

Starting from Eq.~(\ref{gyrokineticequation}) one can follow the same procedure as outlined in Ref.~\cite{pee09c} to 
obtain the equations for the perturbed density normalised to the background density ($\tilde{n}$), and perturbed temperature normalised 
to the background temperature ($\tilde{T}$). For singly charged ions, neglecting the plasma rotation the expressions are   
\begin{equation}
 \omega \tilde{n}  + 2 K \tilde{n} + 2 K \tilde{T} = \langle \tilde{\phi} \rangle \left ( {R\over L_n} - 2 K \right )
\label{deltan} 
\end{equation}
\begin{equation}
 \omega \tilde{T} + \frac{4}{3} K \tilde{n} + \frac{14}{3}K  \tilde{T} = \langle \tilde{\phi} \rangle  \left ({R \over L_T} - \frac{4}{3}K \right )
\label{deltaT}
\end{equation}
where $\omega$ is the frequency normalised to the drift frequency $\omega_D = - k_\theta T / e B R$, 
and $\tilde{\phi}$ here is the perturbed electrostatic potential normalised with $e/T$. Note that all terms that are due to 
the drift are proportional to $K$. We will therefore refer to $\omega_D^* = \omega_D K$ as the effective 
drift frequency. 

The angle brackets in the equation above denote the gyro-average, or FLR effects, which will be modelled 
using a Pade approximation 
\begin{equation}
\langle \tilde{\phi} \rangle = {\tilde{\phi} \over 1 + (k_\perp\rho_i)^2/2}  = F \tilde{\phi},  
\end{equation} 
where $F$ has been introduced to shorten the notation. 
Finally, the gyro-kinetic Poisson equation is solved assuming adiabatic electrons  
\begin{equation}
 \tilde{n} = \left ( 1 + \frac{1}{2}k_\perp^2 \rho_i^2 \right ) \tilde{\phi} = G \tilde{\phi} 
\label{poisson}
\end{equation}
where the term proportional to $k_\perp^2$ is due to the polarization.  

From the equations above, a dispersion relation can be derived
\begin{equation}
 A \left ({\omega \over K} \right )^2 + B {\omega \over K} + C = 0
\end{equation}
where 
\begin{equation}
A = G 
\end{equation} 
\begin{equation}
B =  \frac{20}{3} G + 2 F - {F \over K} {R \over L_N}  
\end{equation} 
\begin{equation}
C = \frac{20}{3} (F + G) + 2{F \over K} {R\over L_T} - \frac{14}{3}{F \over K} {R \over L_N } 
\end{equation}
The growth rate normalised to $\vert \omega_D\vert$ $(\gamma)$ can be readily calculated 
\begin{equation}
\gamma = {\sqrt{2K} \over G} \sqrt{  {R \over L_T} - K {R \over L_{Tcrit}}}    
\label{growthrate}
\end{equation}
where the critical gradient is given by 
\begin{equation}
{R \over L_{Tcrit}} = \frac{1}{8} B^2 - \frac{10}{3}(1 + G^2) + \frac{7}{3}{1 \over K} {R \over L_N}   
\end{equation} 
and we have used $FG = 1$. 

\begin{figure}[h]
 \begin{center}
  \includegraphics[width=9cm]{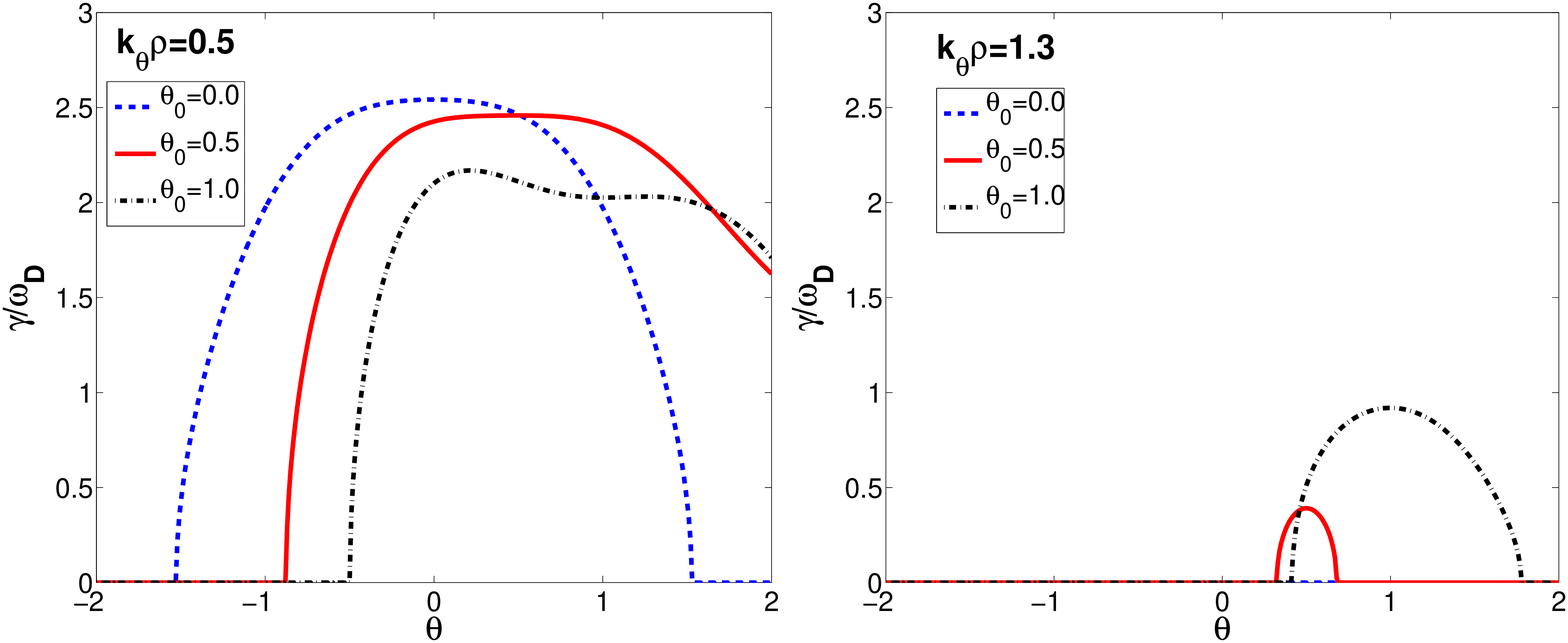}
  \caption{\small (Colour on-line) Growth rates of the fluid model as a function of $\theta$ 
for $\theta_0 = 0$ (dashed), $\theta_0 = 0.5$ (solid) and $\theta_0 = 1.0$ (dash-dotted). The left
panel gives the growth rates for $k_\theta \rho_i = 0.5$, whereas the right panel gives the results
for $k_\theta \rho_i = 1.3$. 
\label{fluidmodel}}
 \end{center}
\end{figure}

Fig.~\ref{fluidmodel} shows the results of the growth rate, normalised to $\vert \omega_D\vert$,
of the fluid model as a function of $\theta$ for three values of $\theta_0 = 0,$ 0.5, 1.0. 
The left panel shows the results for $k_\theta \rho_i = 0.5$, whereas the right panel shows 
the results for $k_\theta \rho_i = 1.3$. It can be seen that for $k_\theta \rho_i = 0.5 $ the mode 
has a maximum growth rate for $\theta = 0$, whereas for $k_\theta \rho_i = 1.3$ the mode 
with $\theta_0 = 0$ is stable and the most unstable mode occurs for $\theta_0 
= 1.0$.  

\begin{figure}[h]
 \begin{center}
  \includegraphics[width=9cm]{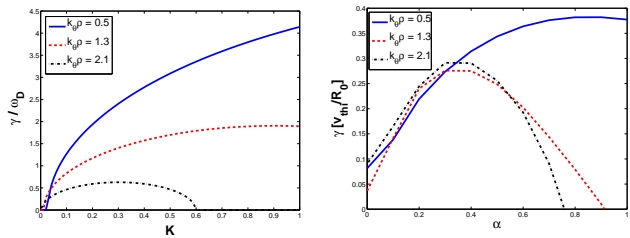}
  \caption{\small (Colour on-line) Left panel: Growth rates (normalised to $\vert \omega_D \vert$) of the 
fluid model as a function of $K$ for $k_\theta \rho_i = 0.5$ full (blue) line $k_\theta \rho_i = 1.3$
dashed (red) line, and $k_\theta \rho_i = 2.1$ dash-dotted black line. Right panel: Growth rates (normalised
to $v_{thi} / R_0$ of spectral gyro-kinetic simulations with $\theta_0 = 0$, with the drift multiplied with 
a constant $\alpha$. The values of $k_\theta \rho_i$ as well as the curve labels 
are the same as that of the fluid model in the left panel. 
\label{Kdep}}
 \end{center}
\end{figure}

The figure of the growth rate shows that the largest growth rate at high $k_\theta \rho_i$ is obtained for 
$\theta \approx \theta_0$, i.e. for $k_r \approx 0$. 
The small value of $k_r$ then does not increase the FLR and polarization stabilisation of the mode. 
Next, we clarify why the mode is strongly stabilised for $\theta_0 =0$ and has its maximum 
growth rate for $\theta_0 \ne 0$. 
For $\theta \approx \theta_0$, $k_r = 0$ and $K = \cos\theta$. 
Therefore $K = 1$ at the low field side position ($\theta = 0$) and decreases for $\theta \ne 0$. 
Eq.~(\ref{growthrate}) gives the dependence of the growth rate on $K$. 
If $K$ is treated as a free parameter, and the density gradient is chosen to be zero for simplicity 
$R/L_N = 0$, then a maximum in the growth rate is obtained for 
\begin{equation} 
K_M = \frac{1}{2}{R / L_T \over  R/L_{Tcrit}}
\end{equation} 
i.e. when $R/L_T > 2 R/L_{Tcrit}$ a maximum growth rate is obtained for the low field side position 
whereas for $R/L_T < 2 R / L_{Tcrit}$ the maximum growth rate will be obtained for $\theta \ne 0$. 
As $k_\theta \rho_i$ is increased for fixed $R/L_T$, $R/L_{Tcrit}$ increases, $K_M$ decreases, and 
the mode shifts away from the low field side. 
The dependence of $\gamma$ on $K$ is shown in Fig.~\ref{Kdep} for various values 
of $k_\theta \rho_i$. It can be seen that for $k_\theta \rho_i = 0.5$ the low field side position 
is the position for which the maximum is reached, while it is shifted away from the low field side 
for $k_\theta \rho_i > 1$. 
 
The physical reason for a maximum in $K$ can be understood as follows. 
The ITG generates ion temperature perturbations due to the perturbed ExB velocity in the background gradients
(the term proportional to $R/L_T$ on the right hand side of Eq.~(\ref{deltaT})). 
Since the drift (${\bf v}_D$) is a function of the particle energy, the temperature perturbations generate 
density perturbations through the convection (the term $2KT$ on the left hand side of Eq.~(\ref{deltan}). 
These ion density perturbations then lead to the generation of the electric field (Eq.~(\ref{poisson}) 
which is responsible for the perturbed ExB velocity.  
For $K = 0$, the convection due to the drift is zero and the mode is stable. 
One might therefore expect that a higher $K$ leads to a more unstable mode, and to some extent this 
is indeed the case, as is clear from Eq.~(\ref{growthrate}) which predicts $\gamma \propto \sqrt{K}$. 
However, the Eqs.~(\ref{deltan},\ref{deltaT}) also contain terms that have a stabilising effect: The change 
in kinetic energy of the ions due to the drift motion in the perturbed potential (the term $-4 K /3\langle \phi \rangle$ on 
the right hand side of Eq.~(\ref{deltaT}), the temperature perturbations that are generated by the 
perturbed density perturbations (the term $4 Kn/3$ on the left hand side of Eq.~(\ref{deltaT})), and 
the fact that density and temperature perturbations have a tendency to propagate with different phase 
velocities. 
These stabilising terms are responsible for the threshold of the mode, and are all proportional to $K$. 
When the threshold is increased by FLR and polarization effects, and is close to $R/L_T$, the largest 
growth rate is obtained for $K < 1$, i.e. a mode shifted away from the low field side. 

The fluid model is, of course, a strong simplification compared with the full gyro-kinetic model. 
The fluid model not only suggests that all instabilities close to the threshold would have their maximum growth rate 
away from the low field side, it also finds no threshold for the ITG, since for 
any finite $R/L_T$, $K$ can be chosen small enough that an instability arises. 
In particular the parallel dynamics (Landau damping) contained in the gyro-kinetic model must be 
considered. This stabilising mechanism is independent of $K$ and can be expected to stabilise any instability 
for which $K$ is too small.  
Nevertheless, if the explanation based on the fluid model is correct, its predictions should be 
qualitatively reproducible by the gyro-kinetic simulations. We discuss two tests below. 

First, we can artificially multiply the drift velocity with a factor $\alpha$ ($0\le \alpha \le 1$), 
in spectral simulations with $\theta_0 = 0$. 
This reduces the drift frequency and is as if we introduce the factor $K$ of the fluid model into the 
gyro-kinetic simulations (with $\alpha = K$). 
For those modes that have a maximum growth rate when the mode is shifted away from the low field side, 
one expects the maximum growth rate for $\theta_0 = 0$ to be obtained for $\alpha < 1$, if the physics 
mechanism discussed above is correct. 
The right panel of Fig.~\ref{Kdep} shows the growth rates of the gyro-kinetic simulations as a function 
of $\alpha$ for the same values of $k_\theta \rho_i$ as the fluid model (shown in the left panel). 
Indeed, the gyro-kinetic simulations at high $k_\theta \rho_i$ are stable for $\alpha = 1$ and have a 
maximum in the growth rate for $\alpha < 1$, qualitatively reproducing the fluid model. 

\begin{figure}[h]
 \begin{center}
  \includegraphics[width=9cm]{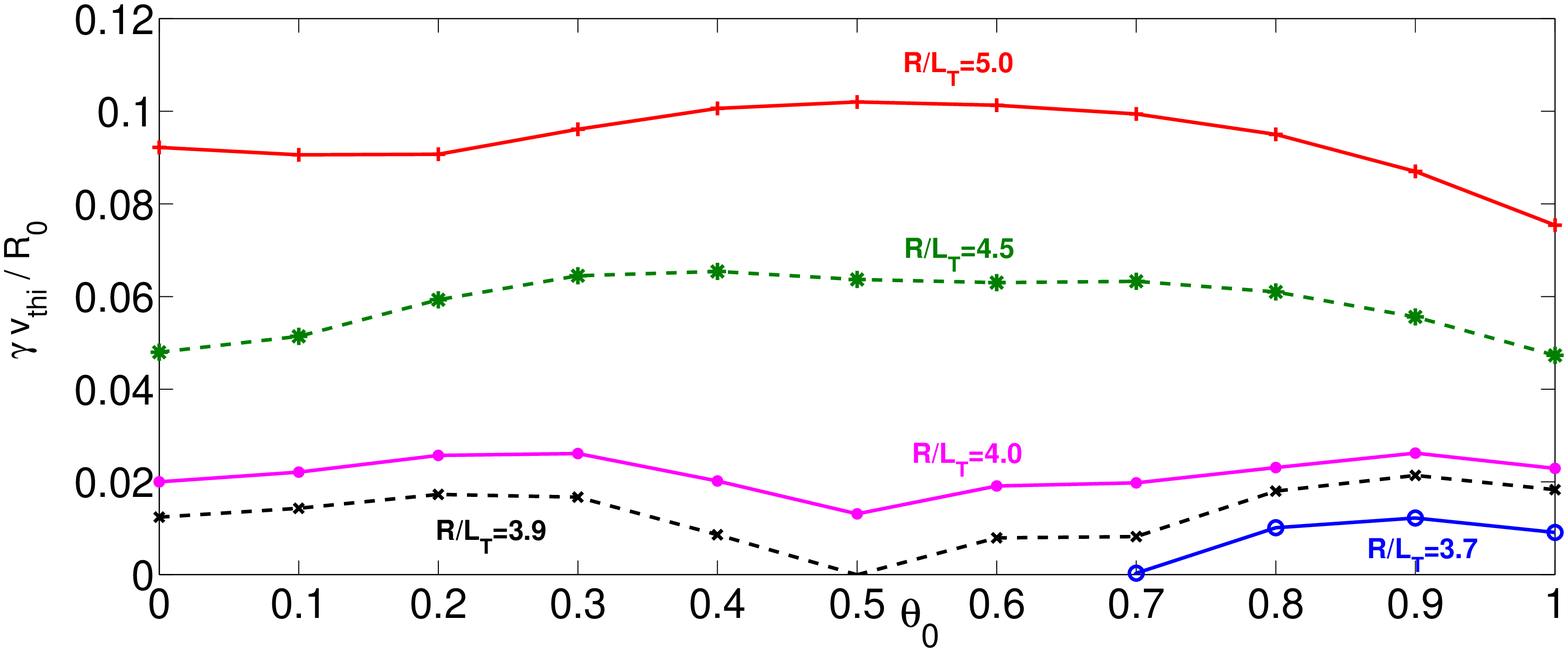}
  \caption{\small (Colour on-line) Growth rates as a function of $\theta_0$ for the Waltz standard 
case with $k_\theta \rho_i = 0.5$, and $R/L_T = 3.7$ open (blue) circles, $R/L_T = 3.9$ (black) 'x', 
$R/L_T = 4.0$ closed (magenta) circles, $R/L_T = 4.5$ (green) stars, and $R/L_T = 5.0$ (red) '+'.   
\label{growthatthreshold}}
 \end{center}
\end{figure}

Second, as discussed above, the mechanism is not limited to $k_\theta \rho_i > 1$. Close to the 
threshold, the most unstable mode can be expected to be shifted away from the low field side (provided
the Landau damping is small enough). 
Fig.~\ref{growthatthreshold} shows the growth rate as a function of $\theta_0$ of the Waltz standard 
case with $k_\theta \rho_i = 0.5$ for several values of $R/L_T$ close to the threshold of the mode. 
Although the effect is small, the largest growth rate is obtained for $\theta_0 \ne 0$. 
In fact, for $R/L_T = 3.7$, an unstable mode exists for $\theta_0 \ne 0$ whereas the mode at $\theta_0 
= 0$ is stable, i.e. a mode shifted away from the low field side exists for a temperature gradient length 
below the threshold of the ITG obtained for $\theta_0 = 0$. 
Both tests give confidence that the physical mechanism found through the analytic fluid model is indeed the 
reason for the observed behaviour of the gyro-kinetic simulations.  

\section{COMPARISON WITH PREVIOUS WORK} 

\begin{figure}[h]
 \begin{center}
  \includegraphics[width=9cm]{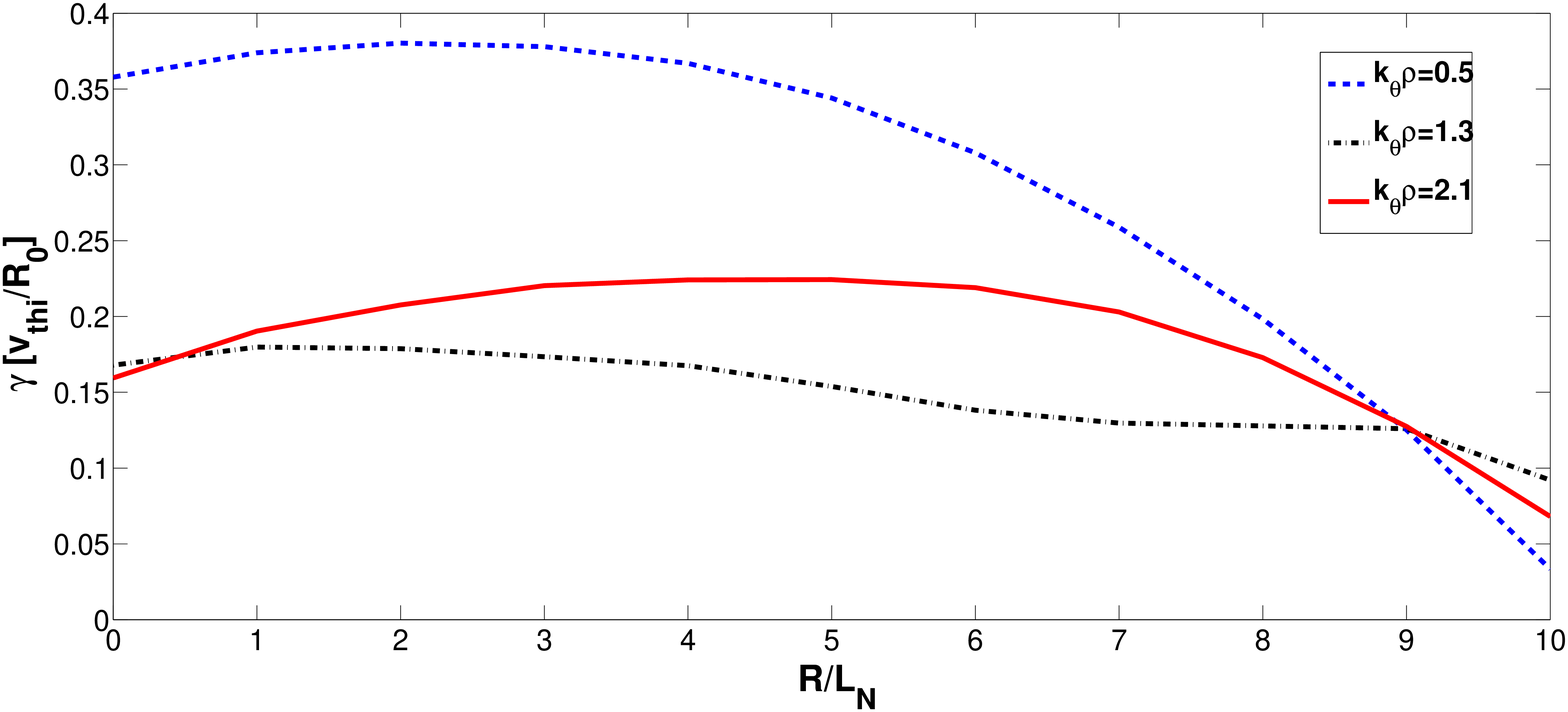}
  \caption{\small (Colour on-line) Growth rates as a function of $R/L_N$ for various values 
of $k_\theta \rho_i$.   
\label{rlnscan}}
 \end{center}
\end{figure}

Ion temperature gradient instability at sub-Larmor radius scales 
have previously been reported in the literature\cite{SMO02,HIR02,GAO03,GAO05,CHO09}. 
These modes have been found in slab geometry, as well as in the case of weak toroidicity. 
The latter condition translates to a density gradient $R /L_N > 6$ for the instability to occur
\cite{HIR02,CHO09}. 
Such a high gradient is not usually obtained in Tokamak plasmas under normal operation. 
In contrast the high $k_\theta \rho_i$ ITG described in this paper occurs for a wide range 
of $R/L_n$ as shown in Fig.~\ref{rlnscan}, and is unstable for $R/L_n = 0$. 

There are similarities between previous work and ours. In Refs.~\cite{SMO02,HIR02} 
it is stressed that the non adiabatic response of the ions at $k_\theta \rho_i > 1$ is 
essential for the instability to occur. A similar statement can be made for the modes discussed
in this paper. However, the essential ingredient discussed in this paper, the shift of the 
mode away from the low field side, reducing the effective drift frequency, 
is a distinct mechanism from that of the works published to date. In particular, an inspection of the equations 
in Refs.~\cite{SMO02,HIR02,GAO05} shows that all these references assume $\theta_0 = 0$.   

\section{CONCLUSION}

In this paper we have shown that
\begin{itemize} 
\item The ITG with adiabatic electrons for standard parameters can be unstable for 
$k_\theta \rho_i$ substantially larger than one. 
\item Essential for this instability is a reduction of the effective drift frequency through 
the shift of the mode away from the low field side. 
\item An enhancement of the growth rate through the reduction of the effective drift frequency 
can be important for $k_\theta \rho_i <1$, in particular close to the threshold.  
\item Unstable modes with $\theta_0 \ne 0$ can exist for ion temperature gradient lengths 
below the threshold of the mode obtained with $\theta_0 = 0$. 
\end{itemize} 
The existence of these modes might set additional requirements on resolution in nonlinear 
runs, and might play a role in small scale zonal flow generation. 

\noindent {\bf Acknowledgement} 

Discussions with R. Singh and S. Brunner are gratefully acknowledged.

\end{document}